\documentclass[twocolumn,showpacs,superscriptaddress,amssymb,10pt,prd,floatfix]{revtex4-1}

\usepackage{graphicx}
\usepackage{dcolumn}
\usepackage{bm}
\usepackage{epsfig}
\usepackage{color}
\usepackage{longtable}
\usepackage{hyperref}
\hypersetup{hidelinks}
\usepackage{enumerate}
\usepackage{ulem}
\usepackage{color}

\newcommand{\p}{\partial} 				
\newcommand{\e}{\mathrm{e}}				
\renewcommand{\i}{i}					
\renewcommand{\d}{\mathrm{d}}			
\renewcommand{\vec}{\bm}				
\newcommand{\Ai}{\mathrm{Ai}}		 	
\newcommand{\F}{\mathcal{F}}   			
\renewcommand{\emph}{\textit}

\definecolor{gray}{rgb}{0.5,0.5,0.5}
\newcommand{\add}[1]{\textcolor{blue}{#1}}
\newcommand{\skipp}[1]{\textcolor{gray}{\sout{#1}}}

\renewcommand{\add}[1]{#1}
\renewcommand{\skipp}[1]{}
\newcommand{\microtab}[1]{#1}

\begin{document}

\preprint{}
%
%
%
%
\title{\skipp{Probing Einstein's theory of gravity} \add{Gravitational light deflection}\\in Earth-based laser cavity experiments}
%
%
%
%

\author{S.~Ulbricht}
\email{sebastian.ulbricht@ptb.de}
\affiliation{Physikalisch--Technische Bundesanstalt, D--38116 Braunschweig, Germany}
\affiliation{Technische Universit\"at Braunschweig, D--38106 Braunschweig, Germany}

\author{J.~Dickmann}
\email{johannes.dickmann@ptb.de}
\affiliation{Physikalisch--Technische Bundesanstalt, D--38116 Braunschweig, Germany}
\affiliation{Technische Universit\"at Braunschweig, D--38106 Braunschweig, Germany}

\author{R.~A.~M\"uller}
\affiliation{Physikalisch--Technische Bundesanstalt, D--38116 Braunschweig, Germany}
\affiliation{Technische Universit\"at Braunschweig, D--38106 Braunschweig, Germany}

\author{S.~Kroker}
\affiliation{Physikalisch--Technische Bundesanstalt, D--38116 Braunschweig, Germany}
\affiliation{Technische Universit\"at Braunschweig, D--38106 Braunschweig, Germany}

\author{A.~Surzhykov}
\affiliation{Physikalisch--Technische Bundesanstalt, D--38116 Braunschweig, Germany}
\affiliation{Technische Universit\"at Braunschweig, D--38106 Braunschweig, Germany}

\date{\today}

%
%
%
%

\begin{abstract} 
As known from Einstein's theory of general relativity, the propagation of light in the presence of a massive object is affected by gravity.
In this work, we discuss whether the effect of gravitational light bending can be observed in Earth-based experiments, using high-finesse optical cavities.
In order to do this, we theoretically investigate the dynamics of electromagnetic waves in the spacetime of a homogeneous gravitational field and give an analytical expression for the resulting modifications to Gaussian beam propagation.  
This theoretical framework is used to calculate the intensity profile at the output of a Fabry-P\'erot cavity and to estimate the imprints of Earth's gravity on the cavity output signal.
In particular, we found that gravity causes an asymmetry of the output intensity profile.
Based on that, we discuss a measurement scheme, that could be realized in facilities like the GEO600 gravitational wave detector and the AEI 10 m detector prototype.

\end{abstract}
\maketitle
\section{Introduction}
\microtab{\vspace{-0.3em}}One century ago, one of the most famous predictions of Einsteins theory of gravitation was approved.
Two observational expeditions led by  A.~S.~Eddington and C.~Davidson observed the stellar light deflection during the solar eclipse of 1919 \cite{Einstein16,Edd19}.
Since then a number of astrophysical observations, i.e. of gravitational lensing, confirmed the gravitational light bending effect \cite{astro1,astro2,astro3}.
In contrast to the variety of astrophysical confirmations, much less is reported about gravitational light deflection in Earth-based experiments \cite{Thorne77,Sivi07,Sivi08,Rich19}.
Today, such experiments likely become possible owing to the advances in high-finesse optical cavities.
These cavities are essential to the nowadays most accurate measurement devices, e.g. optical clocks, high resolution spectroscopy lasers and gravitational wave detectors \cite{Harry2010, Acernese2015, Somiya2012, Kessler2012, Audley2017}. 
Due to the increasing demand for precision in these experiments, the finesse of laser stabilization cavities underwent a tremendous improvement towards  $\F\sim 3\times 10^5$ during the last decades \cite{Ludlow2007, Kessler2012, Matei2017}. 
Moreover, the ongoing research in advanced mirror technologies, like crystalline coatings \cite{Cole2016,Cole2013,Gregory2012} or etalons \cite{Dickmann2018-2,etalons1,etalons2} is promising for further improvements towards even higher finesses.
%
%

\begin{figure}[b!]
	\includegraphics[scale=0.40]{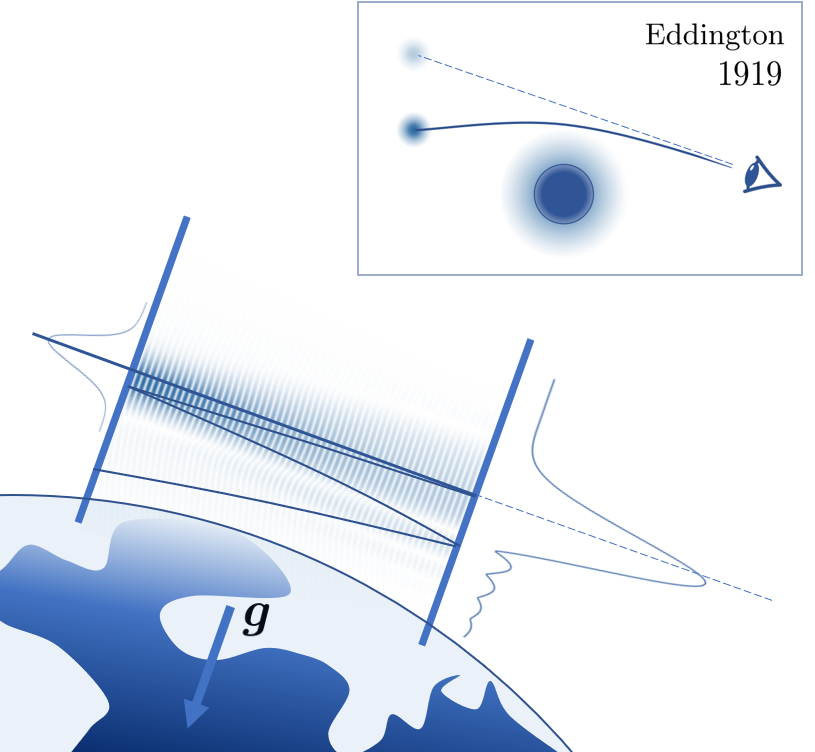}\\	
	\caption{Our proposal for testing gravitational light deflection in a high-finesse laser cavity on Earth. In this experiment the effect of gravity on a Gaussian beam could be detected via changes of the intensity profile at the cavity output. For comparison, the box above shows the gravitational light deflection at the Sun, observed during a solar eclipse, as it was done in 1919 in order to test Einsteins theory of gravitation via displacement of star positions. 
	}
	\label{FIG01}
\end{figure}

In this \skipp{contribution} \add{work}, we propose to use high-finesse cavities to measure gravitational effects.
If such a cavity is subjected to a gravitational field, the light propagation inside the cavity slightly changes and the light beam literally \emph{falls down} while it is caught between two highly reflective mirrors.
This results in an asymmetric cavity output signal, that is expected to be measurable in high-finesse cavities, see Fig.~\ref{FIG01}. 
The fact, that the properties of the underlying structure of spacetime are imprinted on the intensity profile at the cavity output, 
\add{may help to probe the gravitational interaction of light and matter at the laboratory scale.}

\add{The coupling of light to matter is one consequence of the Einstein equivalence principle.
	Recent experiments, like the \emph{Galileo} satellites \emph{Doresa} and \emph{Milena} probe the Einstein equivalence principle by redshift effects on the long range motion of an object \cite{Galileo1,Galileo2}. Unlike that, our proposed experiment could test the coupling of light to gravity at small scales, using the gravitational light bending effect, which exceeds redshift effects by many orders of magnitude.
Moreover, in comparison to the Eddington experiment, where Sun's gravity acts as a scattering potential for the long range propagation of light, in a cavity the effect of light deflection can be investigated directly within the interaction region of the light beam and the gravitational field of our planet.
Therefore, the investigation of cavity internal light bending} may enable a wide range of Earth-based tests of general relativity \skipp{and alternative theories of gravitation} at the laboratory scale \add{in a controlled and reproducible experimental environment.}\microtab{\vspace{-0.3em}}

%
\vspace{0,5em}
\section{Light propagation in a homogeneous gravitational field} \label{sec:maxwell}
\vspace{0,5em}
In what follows, we will build up a formalism, that allows us to investigate the influence of gravity on a laser beam with Gaussian intensity profile, that enters a horizontal Fabry-P\'erot cavity.
Before we discuss this scenario, however, we have to develop the formalism from the basic principles of light propagation in a gravitational field. 
Within the framework of classical electromagnetism, the propagation of light is governed by Maxwell equations. 
This set of differential equations for the electric and magnetic fields $\vec{E}$ and $\vec{B}$, can be rewritten in terms of the vector and scalar potentials $\vec{A}$ and $\Phi$. 
Considering Lorenz gauge condition, in the case of freely propagating light, the scalar potential is determined by the vector potential, from which the fields  $\vec{E}=-\p_t\vec{A}-\nabla \Phi[\vec{A}]$ and $\vec{B}=\nabla\times \vec{A}$ can be obtained. 
This way, the set of Maxwell equations reduces further to a wave equation that in the flat spacetime of an inertial observer reads $\p_t^2\vec{A}/c^2-\nabla^2\vec{A} =0$.
Here $c$ is the speed of light, while $\p_t$ is the derivative with respect to time and $\nabla$ is the nabla operator. 
%
%

In the presence of a gravitational field, however, the properties of spacetime change, which leads to modifications of the wave equation \cite{Wald}. In the case of the approximately homogeneous gravitational field of the Earth, characterized by the acceleration $g=-9.81\,\mathrm{m}/\mathrm{s}^2$ \cite{Remark3,Rind60,Rind66} this modified wave equation reads
\begin{equation}
\frac{1}{c^2}\frac{\p^2}{\p t^2}\vec{A} -\vec{\mathcal{D}}^2\vec{A}=0 + \mathcal{O}\left(\epsilon^2\right)\,, \label{eqn:wave}
\end{equation}
where $\vec{\mathcal{D}}=\left(1-\vec{g}\cdot\vec{r}/c^2\right)\!\!\nabla$ differs from the usual nabla operator $\nabla$ by a factor, that accounts for the \emph{gravitational light deflection} and \emph{redshift}. 
Here, we have considered effects of gravity only to linear order in the dimensionless parameter $\epsilon = gL/c^2$, which is in the range of $\epsilon\sim 10^{-18}\dots 10^{-13}$ for typical length scales $L\sim1\,\mathrm{cm}\dots 1\,\mathrm{km}$ of the experiment.
Moreover, $L$ has to be small in comparison to the Earth's radius, such that the gravitational field can be considered as homogeneous.
%
%

The modified wave equation (\ref{eqn:wave}) is a linear partial differential equation for the vector potential $\vec{A}$. This means that a superposition of solutions to (\ref{eqn:wave}) also solves the equation.
Under the assumption of quasi scalar beam propagation, therefore, any solution of the wave equation can be constructed by a superposition of the scalar basis functions
\begin{eqnarray}
\psi^{\vec{k}}_{\delta k_z}(t,\vec{r})= \frac{1}{\,2\pi \sqrt{\delta k_z}}\ \e^{\frac{g z}{2c^2}}\,\Ai\left[-\frac{k_z^2}{\delta k_z^2}- \delta k_z z \right]\label{eqn:function}\\
\times\, \e^{i k_x x} \e^{i k_y y} \e^{-i \omega_0 t} \,,\nonumber
\end{eqnarray}
that solve Eq.~(\ref{eqn:wave}) for a specific frequency constant $\omega_0$ and a constant wave vector $\vec{k}=(k_x,k_y,k_z)$, which are connected via the dispersion relation $c^2\vec{k}^2=\omega_0^2$ \cite{Remark1}. 
We find that the behavior of the wave in $z$-direction, which is chosen along the acceleration vector $\vec{g}=(0,0,g)$, is described by a damped Airy $\mathrm{Ai}$-function (cf. \cite{Sivi07,Sivi08}). The gravitational effects enter this expression via the damping term $g/2c^2$ and the quantity $\delta k_z  =  (2 g \omega_0^2/c^4)^{1/3}$.
In the case $g=0$, the basis functions (\ref{eqn:function}) resemble the well-known plane wave solutions $\psi^{\vec{k}}_{0}(t,\vec{r})\sim \e^{\i( \vec{k}\cdot\vec{r}-\omega_0 t)}$, which describe light propagation in the spacetime of an inertial observer. 
%

\section{Gravitational effect on Gaussian beam propagation}
\begin{figure}[t]
	\includegraphics[scale=0.42]{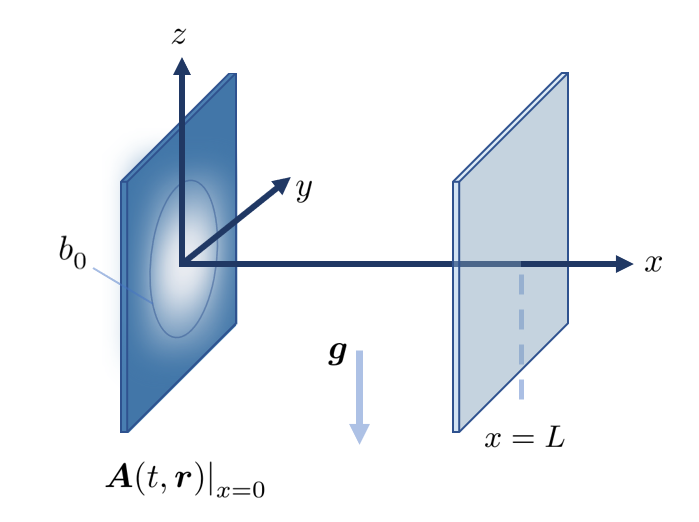}
	\caption{Choice of coordinates in the Fabry-P\'erot cavity setting, consisting of two plane mirrors. The coordinate center is situated at the maximum of the initial Gaussian intensity profile at the first mirror, located in the $(y,z)$-plane. The second mirror is placed at $x=L$.
		This geometry sets the boundary conditions for the propagation of a gravitationally modified Gaussian beam and the calculation of the cavity output signal.}
	\label{FIG02}
\end{figure}
With the complete set of basis functions (\ref{eqn:function}), we are able to model light propagation in a homogeneous gravitational field for any boundary condition \cite{Hunt81}.
In this \skipp{Letter} \add{paper} in particular, we consider a typical experimental setup, where a $y$-polarized laser beam with Gaussian intensity profile enters a Fabry-P\'erot cavity. For this setup, the vector potential at the input mirror, located at $x=0$, is given by
\begin{equation}
\left.\vec{A}(t,\vec{r})\right|_{x=0}=\frac{\vec{A}_0}{2\pi b_0^2}\,\e^{-\frac{z^2+y^2}{2b_0^2}}\e^{-\i \omega_0 t}\,, \label{boundary}
\end{equation}
where $\vec{A}_0$ is a constant vector pointing into $y$-direction and $b_0$ is the waist of the beam, see Fig. \ref{FIG02}. 
For our further analysis, it is convenient to expand (\ref{boundary}) in terms of the basis functions (\ref{eqn:function}), see \cite{Vallee}.
This expansion reads
$\left.\vec{A}(t,\vec{r})\right|_{x=0}=\left(1-gz/2c^2\right)\int\tilde{\vec{A}}^b_{\vec{k}}\psi^{\vec{k}}_{\delta k_z}(t,\vec{r})\,\,\d k_x \d k_y \d (k_z^2)$, where the coefficients 
\begin{eqnarray}
\tilde{\vec{A}}^b_{\vec{k}}=\vec{A}_0 \exp\left[{-\frac{1}{2}(k_z^2+k_y^2)b_0^2}\right]\delta(k_x)\hspace{4em}\label{eqn:Ak}\\
\times\frac{1}{\sqrt{\delta k_z}}\Ai\left[-\frac{k_z^2}{\delta k_z^2}+\frac{1}{4}(\delta k_z b_0)^4\right]\,,\nonumber
\end{eqnarray}
are found by projecting the boundary condition $(\ref{boundary})$ to the complex conjugated basis functions $\psi^{\vec{k}\ast}_{\delta k_z}(t,\vec{r})$.
The first line of this expression is the Fourier transformation of the Gaussian profile (\ref{boundary}).
In the presence of gravity, however, $\tilde{\vec{A}}^b_{\vec{k}}$ also contains an Airy $\mathrm{Ai}$-function in the second line, which becomes unity for $g=0$, such that we recover the result of standard Fourier optics in the case of vanishing gravity.
\begin{figure}[t]
	\vspace{0.3em}	\hspace{1.1em}(a)\hspace{11.1em} (b)\hfill $\left.\right.$\\[1em]
	\includegraphics[scale=0.33]{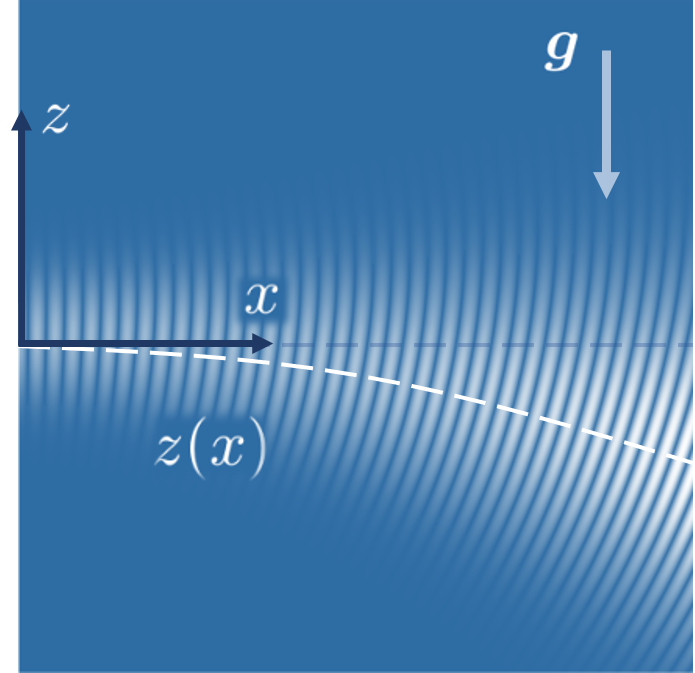}\,\,\, \includegraphics[scale=0.33]{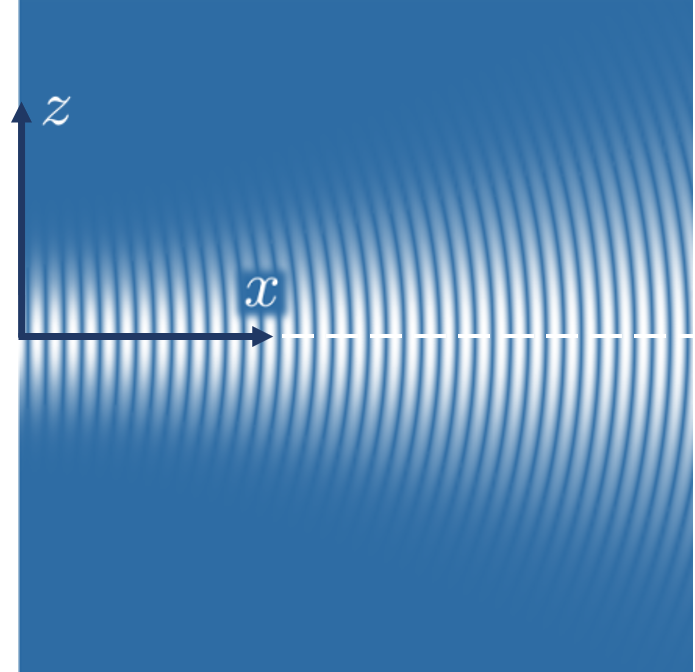}
	\caption{Schematic illustration of the Gaussian beam propagation (a) with and (b) without gravitational modifications. In the presence of gravity, the intensity maximum follows the parabola $z(x)=-|g|x^2/2c^2$.\vspace{-1em}}
	\label{FIG03}
\end{figure}
%
%

The coefficients $\tilde{\vec{A}}^b_{\vec{k}}$ from Eq.~(\ref{eqn:Ak}) only describe the beam at $x=0$, which is encoded in the delta function $\delta(k_x)$. However, we want the beam to obey the wave equation (\ref{eqn:wave}) for all points in space, close to the beam axis. Therefore, we implement the dispersion relation $c^2\vec{k}^2=\omega_0^2$ by the replacement of the delta function $\delta(k_x) \to \delta(k_x - \omega_0/c+ c(k_y^2+k_z^2)/2\omega_0)$. At this point, the approximation of paraxial light propagation along the $x$-axis, with the assumption $k_y,k_z\ll k_x$, is taken into account. The new coefficients, denoted as $\tilde{\vec{A}}_{\vec{k}}$, can be employed to calculate the vector potential of the gravitationally modified Gaussian beam via $\vec{A}(t,\vec{r})=\left(1-gz/2c^2\right)\int\tilde{\vec{A}}_{\vec{k}}\psi^{\vec{k}}_{\delta k_z}(t,\vec{r})\,\,\d k_x \d k_y \d (k_z^2)$. The vector potential can be further evaluated by applying methods as presented in \cite{Vallee}. In result, the beam can be decomposed into
\begin{eqnarray}
\vec{A}(t,\vec{r})\,\,
&=&\vec{A}^{\mathrm{G}}(t,\vec{r})\, \e^{S_g(\vec{r})}+ \mathcal{O}\left(\epsilon^2\right)\,.\label{beam}
\end{eqnarray}
Here $\vec{A}^{\mathrm{G}}(t,\vec{r})$ describes the propagation of a usual Gaussian beam in the absence of gravity \cite{Remark2},\microtab{\vspace{1em}} while its gravitational modifications are collected in the complex exponent
\begin{eqnarray}
S_g(\vec{r})=\frac{ g\omega_0^2 b_0^2 z}{2c^4(1+\mu^2)}\left[\mu^2+\i (2 \mu+ \mu^3)\right]\, ,\label{eqn:Sg}
\end{eqnarray}
where we introduced $\mu=x/x_R$, which is the length of propagation in units of the Rayleigh length $x_R=b_0^2\omega_0/c$. The real part of $S_g(\vec{r})$ describes the falling of the intensity profile, 
such that the intensity maximum of the beam follows the parabola $z(x)=-|g|x^2/2c^2$, see Fig.~\ref{FIG03}a. 
Comparing  the real part of $S_g(\vec{r})\sim (\omega_0 b_0/c)^2 (g z /c^2)$ with redshift contributions $\sim (g z /c^2)$, we find, that the effect of gravitational light bending, is larger by a factor $(\omega_0 b_0/c)^2\gg1$  and, therefore, exceeds redshift effects by many orders of magnitude.
The imaginary part of $S_g(\vec{r})$ gives rise to a $z$-dependent gravitational phase shift $\phi_g(x\gg x_R)\sim g\omega_0/c^3\,\, z \,x $, that grows while the beam propagates. 
As we will discuss hereafter, the falling of the intensity maximum and the gravitational phase shift will have an measurable effect on the output signal of high-finesse Fabry-P\'erot cavities.
\begin{figure}[b]
	\includegraphics[scale=0.27]{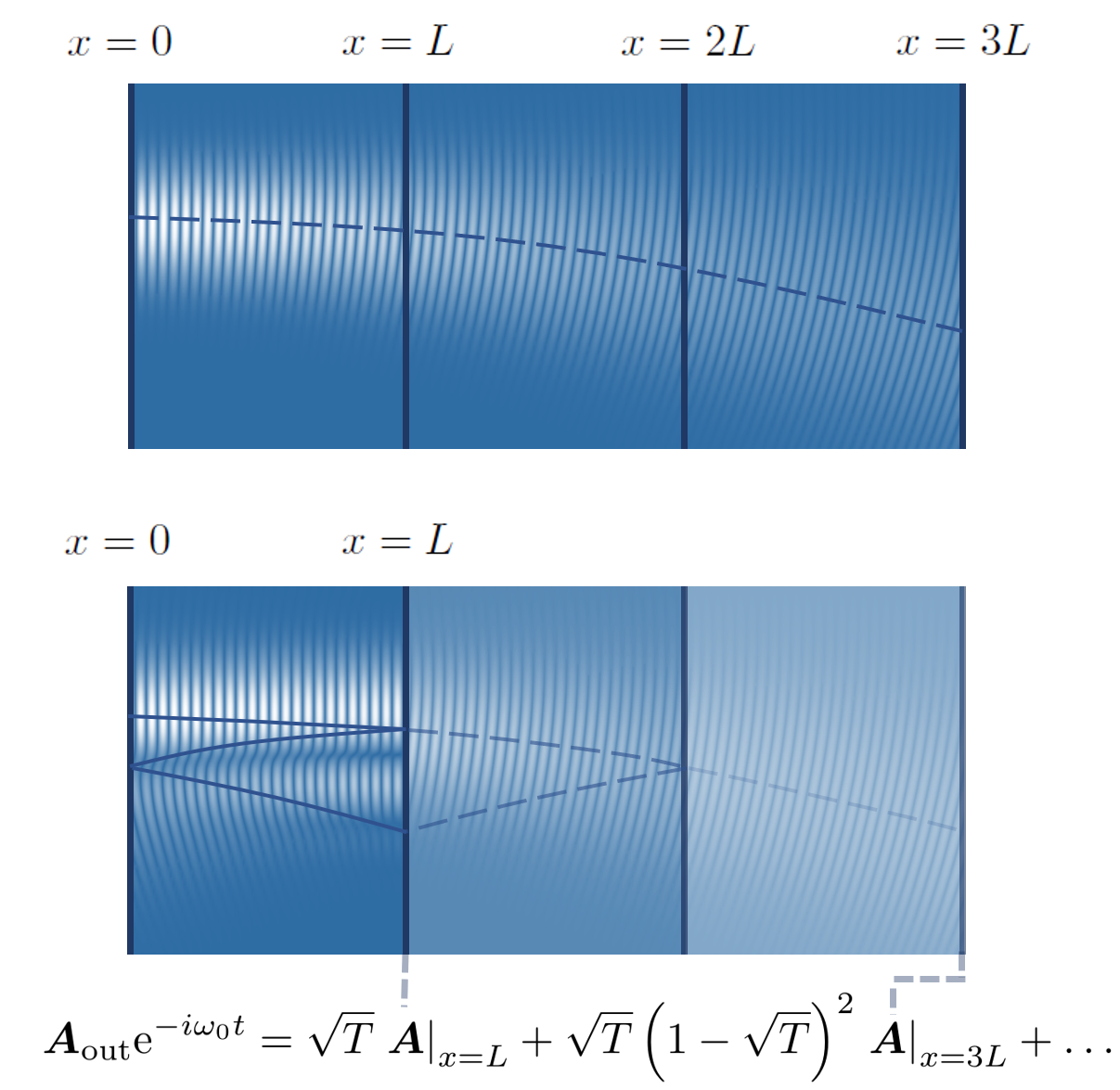}\hspace{3.5em}\\[-0.7em]
	\caption{Illustration of the first round trip of the modified Gaussian beam in a Fabry-P\'erot cavity. The properties of the plane mirrors allow us to model the light distribution in the cavity and its output signal by the overlay of slices of the beam, where every slice is weighted by a function of the transmittance $T$.} \label{FIG04}
\end{figure}
%
%
%

\section{Output signal of a Fabry-P\'erot cavity}

Previously we have shown, that a homogeneous gravitational field modifies the propagation of a Gaussian beam. 
Since the deviations from a usual Gaussian beam are small, however, a long propagation distance would be needed to measure the effect under real experimental conditions. In order to overcome this in a laboratory scale experiment, a Fabry-P\'erot cavity can be used to accumulate the effect.
To analyze this scenario, we assume a cavity, that consists of two identical plane parallel mirrors at the positions $x=0$ and $x=L$, as shown in Fig.~\ref{FIG02}.
Commonly, a Fabry-P\'erot cavity is characterized by its finesse $\mathcal{F}$, that is related to the transmittance of the mirrors via $\mathcal{F}=\pi \sqrt{1-T}/T$ \cite{Ismail2016}, where $T$ determines the number of roundtrips in the cavity. 

The two planar mirrors only change the direction of propagation along the $x$-axis.
This for example means, that the signal at the position of the output mirror after the first two reflections equals the freely propagating beam at the position $x=3L$, as illustrated in Fig.~\ref{FIG04}.
Therefore, the vector potential behind the cavity output mirror can be obtained by the overlay of slices of the modified Gaussian beam 
\begin{eqnarray}
\vec{A}_{\mathrm{out}}(y,z)\,\e^{-\i \omega_0 t}=\label{eqn:output}\hspace{14em}\\
\sum\limits_{n=0}^{\infty}\sqrt{T}\left(1-\sqrt{T}\right)^{2n}\left.\vec{A}(t,\vec{r})\right|_{x=(2n+1)L}\,, \nonumber
\end{eqnarray}
where every slice is weighted by the loss per reflection $(1-\sqrt{T}\,)$ \emph{in} the cavity.
%
%

\begin{figure}[t]
	\hspace{0.7em}(a)\hspace{11.6em} (b)\hfill $\left.\right.$\\[1em]
	\includegraphics[scale=0.33]{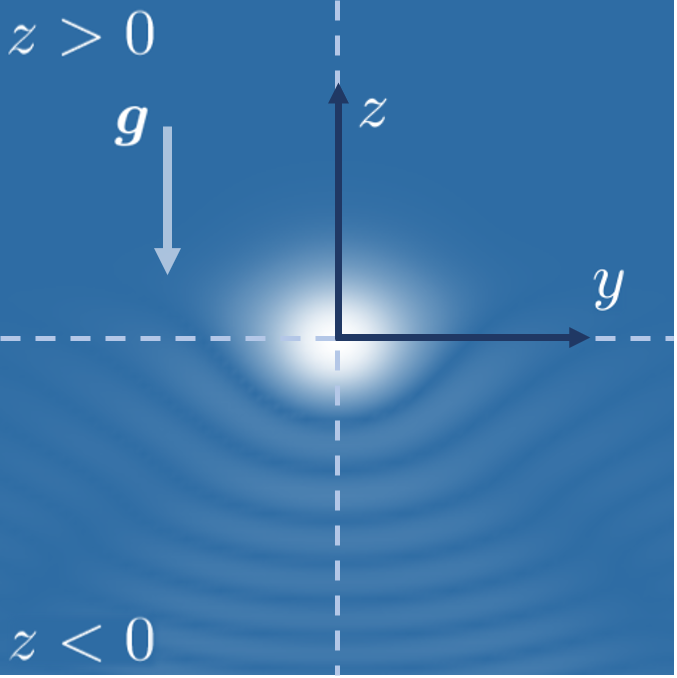}\,\,\,\,\,\,\, \includegraphics[scale=0.33]{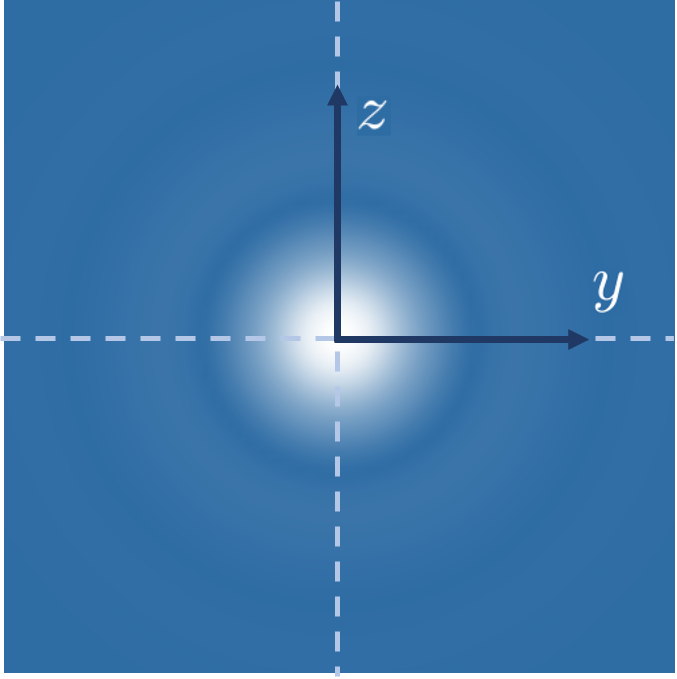}
	\caption{Schematics of the characteristic output signal $|\vec{A}_{\mathrm{out}}(y,z)|^2$ of a Fabry-P\'erot cavity (a) with a highly exaggerated value of $g$ and (b) without gravity. 
		While for the Earth's actual value of $g$ the effect can not be imaged directly in cavities with current precision, a quadrant detector can be used to study the effect by comparison of the total output power in the upper and the lower half-plane. }
	\label{FIG05}
\end{figure}
%
%

With the vector potential (\ref{eqn:output}), we can analyze the properties of the light, that emerges from the Fabry-P\'erot cavity. The intensity profile at the cavity output, in paraxial approximation is given by the energy density at the output mirror, that can be expressed by the squared module of the vector potential $I(y,z)=c\varepsilon_0\omega^2_0(1-gz/c^2)^{-1}|\vec{A}_{\mathrm{out}}(y,z)|^2$ \cite{Remark4}. In fact, the redshift-prefactor of the intensity turns out to be negligible in comparison to the light deflection of the beam that constitutes the leading effect on the structure of the intensity pattern, illustrated in Fig.~\ref{FIG05}. Furthermore, integrating the intensity between two heights $z_1$ and $z_2$ gives the power $P(z_1<z <z_2)=\int_{z_1}^{z_2}\int_{-\infty}^\infty I(y,z)\, \d y\d z$
at the corresponding section of the output mirror. The total output power $P$, therefore, is obtained by integrating the intensity over the whole mirror surface. 
%

\section{Measurement scheme} 
In order to extract the effect of gravity from the signal, a position sensing detector, i.e. a quadrant detector, can be used.
Such a detector measures the power difference between the upper and the lower half-plane of the output mirror.
This difference, that can be normalized to the total output power, determines the detection parameter
\begin{equation}
\chi= \bigl( P(z<0)-P(z>0)\,\bigr)/P\,. \label{eqn:detector}
\end{equation}
\skipp{State of the art shot-noise limited quadrant detectors are capable to measure the parameter $\chi$, up to parts per billion (ppb $\sim 10^{-9}$), depending on the total output power \cite{Edelstein1978}:}
\add{The measurement of this parameter is limited by the detector photon shot noise, that determines the sensitivity $\Delta \chi$ as a function of the total output power $P$ by}
\begin{eqnarray}
\Delta \chi=\sqrt{\left\langle (\Delta P/P)^2 \right\rangle} &=& 2\sqrt{\frac{hc\,\Delta f}{\lambda P}} \label{eqn:shotnoise} \\ 
&\approx& 0.86\times 10^{-9} (P/W)^{-1/2}\, \nonumber,
\end{eqnarray}
where $\left\langle (\Delta P/P)^2 \right\rangle$ is the relative light power variance and $\Delta f$ is the measurement bandwidth, which we set to the common value of $1 \mathrm{Hz}$ \cite{Edelstein1978}.
\add{For cavity lengths of cm to m scale, the detector photon shot noise is the dominant limitation of sensitivity in the proposed experiment, assuming optimized measurement frequencies and experimental parameters such as mirror mass, seismic decoupling and optical coatings \cite{Robinson2019,Dooley2015,Notcutt06}.
For larger cavities, like the GEO600 facility, with increased finesse, cryogenic mirrors or meta-mirrors would be needed to suppress the mirror coating thermal noise below the photon shot noise. \cite{Cole2016,Dickmann2018-2}.}
%
%

In what follows, we  will use the  signal-to-noise ratio  $\chi/\Delta \chi>1$ as a criterion for the measurability of the cavity internal gravitational light bending effect.
In order to obtain the predicted signal of the quadrant detector, the summation in Eq. (\ref{eqn:output}) and the power integrals in Eq. (\ref{eqn:detector}) are evaluated numerically. 

%

\section{Results and discussion} 
The equations (\ref{eqn:output}) - (\ref{eqn:shotnoise}) allow us to analyze the gravitational light bending effect for a wide range of experiments with Fabry-P\'erot cavities.
Therefore, in this \skipp{Letter} \add{rapid communication} we consider three specific showcase scenarios for Earth-based cavity designs:  
(i) a horizontal $21\,\mathrm{cm}$-Fabry-P\'erot cavity with a finesse between $3\times 10^5$ and $3\times 10^7$, (ii)  a hypothetical Fabry-P\'erot version of the GEO600 facility with an improved finesse between $3\times 10^3$ and $3 \times 10^5$ and (iii) the 10 m gravitational wave detector prototype at the Albert-Einstein institute Hannover (AEI) with the same finesse values as for GEO600. Our results are presented in the Table~\ref{Tab1}, where we give the parameter $\chi$ and the ratio $\chi/\Delta \chi$ for each scenario. Moreover, the inverse signal-to-noise ratio can be interpreted as the relative accuracy of hypothetical measurements of the gravitational acceleration $\eta=\Delta g/g$ using the light bending effect. Combining Eq.  (\ref{eqn:detector}) and (\ref{eqn:shotnoise}), the the order of magnitude of the effect can be estimated 
\begin{equation}
\frac{\Delta g}{g}=\frac{\Delta \chi}{\chi}=\frac{c^2}{g b_0}\sqrt{\frac{hc\Delta f}{\lambda P}}\left(\frac{\omega_0b_0}{c}\right)^{-2}s(\sigma,\mathcal{F})\,\,,
\end{equation}
where, in order to get the exact value, the factor $s(\sigma,\mathcal{F})$ depending on the cavity length in units of the Rayleigh length $\sigma=Lc/b^2_0\omega_0$ and the finesse $\mathcal{F}$ has to be evaluated. In what follows, we will discuss the the parameter $\chi$ and the signal-to-noise ratio $\chi/\Delta \chi$ for the cavity settings mentioned above.
%
%

%
%
\begin{table}[t!] \microtab{\vspace{-0.77em}}
	\caption{Results for the parameter $\chi$ and the ratio $\chi/\Delta \chi$ for the cavity design examples of a 21 cm cavity (i), GEO600 (ii) and the AEI 10 m prototype (iii).
		The results  are presented  for recent and hypothetical higher finesses $\F$ of these devices. For the calculations we assumed the cavitiy to be matched to a wavelength of $\lambda=1064\,\mathrm{nm}$.\\[0.5 em]}
	\begin{tabular}{p{2em}p{0.0em}p{7em}|p{0.2em}p{4.2em}p{5em}p{5em}}
		& & Properties			& & $\F$ 				&  $\chi$\, [ppb]	& $\chi/\Delta \chi$ 	\\[0.5em]\hline
		& &						& & 					&					&									\\[0.3em]
		(i)		& & $ L=0.21$ m			& & $3\times 10^{5}$ 	& $2\times 10^{-3}$	& $5\times 10^{-7}$					\\
		& & $ b_0 =0.5$  mm  	& & $3\times 10^{6}$	& $5\times 10^{-3}$	& $1\times 10^{-6}$					\\
		& & $ P =40$ nW			& & $3\times 10^{7}$	& $9\times 10^{-3}$	& $2\times 10^{-6}$					\\
		& &						& &						& 					&									\\
		(ii)	& & $ L=600.00$ m		& & $3\times 10^{3}$	& $\phantom{000}42.2$	&	$\phantom{000}87.4$				\\
		& & $ b_0 =18$ mm		& & $3\times 10^{4}$	& $\phantom{000}89.3$	&	$\phantom{00}184.8$				\\
		& & $ P = 3.2$ W 		& & $3\times 10^{5}$	& $\phantom{00}178.1$			&	$\phantom{00}368.7$				\\
		& &						& &						& 					&									\\
		(iii)	& & $ L=10.00$ m		& & $3\times 10^{3}$	& $\phantom{0000}0.4$	&	$\phantom{0000}2.8$					\\
		& & $ b_0 =7.0$ mm 		& & $3\times 10^{4}$	& $\phantom{0000}1.2$	&	$\phantom{0000}8.6$					\\
		& & $ P = 38$ W  		& &	$3\times 10^{5}$	& $\phantom{0000}2.7$	&	$\phantom{000}19.3$							\\		
	\end{tabular}
	\label{Tab1}
\end{table}
First, we consider the 21 cm-Fabry-P\'erot cavity. Due to the combination of the small beam waist of $0.5$ mm \cite{Matei2017}, low laser power in the nano watt range, e.g $40$ nW \cite{Robinson2019}, and the short length of the cavity, even for higher finesses of $3\times 10^{7}$ no cavity internal gravitational light bending effect can be measured in this setup.
This can be seen from Table \ref{Tab1}, where the quadrant detector signal is less than $10^{-4}$ ppb, which is much smaller than the detector noise.
However, gravity may set a limit on the phase stability of comparable laser stabilization cavities, as we will discuss in a future publication \cite{Ulbricht20xx}.
%
%

The effect can be enhanced dramatically if one uses longer cavities and larger beam waists.
To show this, we discuss a hypothetical Fabry-P\'erot version of the GEO600 facility with a 600 m base line: The substitution of the GEO600 beam splitter by a mirror, which is comparable to the current GEO600 mirrors \cite{Grote2008}, would transform the Michelson interferometer into a Fabry-P\'erot cavity.
This configuration could be operated with a laser beam waist of $18$ mm \cite{Heinert2014}. 
\skipp{Considering finesses up to  $3\times 10^5$ and keeping the current GEO600 input power of 3.2 W \cite{Grote2008}, the signal is expected to be 178 ppb with a signal-to-noise ratio of up to 400 at a measurement bandwidth of 1 Hz.}
\add{Considering a finesse of $3\times 10^3$ and keeping the current GEO600 input power of 3.2 W \cite{Grote2008}, the signal is expected to be 42 ppb with a signal-to-noise ratio of up to 90 at a measurement bandwidth of 1 Hz.
While increasing the finesse up to $3\times 10^5$ could enhance the signal to 178 ppb with a signal-to-noise ratio of up to 400, cryogenic mirrors or meta-mirrors would be needed to reach this sensitivity. Moreover,} by assuming the planned GEO-HF update \cite{Dooley2015} and the associated increased laser power by a factor of 4, the signal-to-noise ratio would be enhanced further by a factor of 2. 
%
%

\microtab{\vspace{3em}}
The needed size of the experiment scales down, if a higher laser power is used. As a third example we, therefore, discuss the AEI 10 m prototype, that can be operated with a powerful 38 W input laser \cite{Westphal2012}. Like in the case of GEO600, the substitution of the beam splitter by a highly reflective mirror would turn the device into a Fabry-P\'erot interferometer.
Considering a beam waist of $7$ mm \cite{bw3} and a finesse of $3\times 10^5$, the detector output signal $\chi$ is expected to be 2.7 ppb with a signal-to-noise ratio of 20 at a 1 Hz bandwidth. Thus, a Fabry-P\'erot version of the AEI 10 m prototype would be suitable for testing the cavity internal gravitational light bending effect \skipp{on the laboratory scale} \add{at laboratory scales}.
%

\section{Summary} 
In this \skipp{Letter} \add{work}, we lay down a theory for the propagation of light in a homogeneous gravitational field. 
For this, we solved the wave equation in Rindler spacetime to first order in $gL/c^2$ and applied the result to the propagation of a Gaussian beam.
We found, that the presence of gravity leads to a falling of the beam intensity profile and an additional phase shift.
Both effects increase while the beam propagates.
%
%

In order to study the gravitational light bending, we propose to perform measurements in high-finesse Fabry-P\'erot cavities, where the effect accumulates due to the multiple roundtrips of the light between the mirrors. 
For such a setup, we applied our theory to  calculate the intensity profile at the cavity output.
In order to estimate whether gravitational effects can be observed in the output signal of nowadays or future cavity settings, we considered three experimental scenarios.
Based on this we found, that cavity designs like the GEO600 gravitational wave detector and the 10 m detector prototype at Albert-Einstein-institute Hannover are highly suitable for the measurement of the gravitational light bending effect.
\add{Moreover, we found that the effect of light bending for a Gaussian beam is enhanced by a factor $(\omega_0b_0/c)^2\gg 1$ in comparison to redshift effects.}
We suppose, that the realization of \skipp{these} \add{the proposed} experiment and the validation of the effect \skipp{could} \add{can} open a road to to test \skipp{well-established and alternative theories of gravity} \add{the interaction of light and gravity at small scales} in Earth-based cavity experiments.
%
%

\begin{acknowledgements}The authors acknowledge the support by the Deutsche Forschungsgemeinschaft (DFG, German Research Foundation) under Germany’s Excellence Strategy - EXC-2123 QuantumFrontiers - 390837967.\\ \vspace{5em}\end{acknowledgements}

%


\begin{thebibliography}{12}
\bibitem{Remark3} We use the metric in its spatially isotropic form with the line element $\d s^2 = \left(1-\vec{g}\cdot\vec{r}/c^2\right)^2\!\!\d(ct)^2-\d \vec{r}^2$.
\bibitem{Remark1} Here we allow for real and imaginary valued $k_z$, what also includes damping in $z$-direction and ensures the completeness of the basis functions with respect to
	$\int \psi^{\vec{k}\ast}_{\delta k_z}(t,\vec{r}')\psi^{\vec{k}}_{\delta k_z}(t,\vec{r})\,\d k_x \d k_y \d(k_z^2)=(1+g z/c^2)\,\delta^{(3)}(\vec{r}-\vec{r}')+ \mathcal{O}\left(\epsilon^2\right)$. Due to the asymptotic properties of the Airy $\mathrm{Ai}$-function, negative valued $k_z^2$ become irrelevant in the limiting case of vanishing $g$ and the standard Fourier basis is recovered.
\bibitem{Remark4}
		The energy density of the gravitational affected Gaussian beam reads 
		$w(\vec{r})=\frac{\varepsilon_0}{2}\left(1-\frac{gz}{c^2}\right)^{-1}|\p_t\vec{A}+\nabla \Phi[\vec{A}]|^2+\frac{1}{2\mu_0}\left(1-\frac{gz}{c^2}\right)|\nabla\times\vec{A}|^2$.
		The scalar potential $\Phi[\vec{A}]=-\frac{\i\lambda}{2\pi b_0}\left(1-\frac{gz}{c^2}\right)^2\frac{\vec{r}\cdot \vec{A}(t,\vec{r})}{b_0(1+\i\mu)}$ is determined by the vector potential via Lorenz gauge condition. Thus, the contributions of $\Phi[\vec{A}]$ to the energy density are by a factor $(\lambda/2\pi b_0)^2$ smaller than contributions of the vector potential: 	$w(\vec{r})=\varepsilon_0\omega^2_0(1-\frac{gz}{c^2})^{-1}\left(1-(1-\frac{gz}{c^2})^{2}\frac{\lambda^2}{(2\pi b_0)^2}\frac{y^2}{2b_0^2(1+\mu^2)}\,\right)|\vec{A}|^2$.
		The contribution of $\Phi[\vec{A}]$ vanishes at the beam axis and shrinks while the beam propagates. In result the scalar potential does not affect the outcome of the calculations quantitatively.
\end{thebibliography}
\end{document}